\documentstyle[prd,preprint,tighten,aps,eqsecnum,%
amssymb,newlfont]{revtex}
\begin{document}
\preprint{
\begin{tabular}{r}
UWThPh-2001-1\\
January 2001
\end{tabular}
}
\draft
\title{A model for decoherence of entangled beauty\thanks{This
research was performed within the FWF Project No.\ P14143--PHY of the
Austrian Science Foundation.}}
\author{R.A. Bertlmann and W. Grimus}
\address{Institut f\"ur Theoretische Physik, Universit\"at Wien\\
Boltzmanngasse 5, A-1090 Vienna, Austria}

\maketitle

\begin{abstract}
In the context of the entangled $B^0 \bar B^0$ state 
produced at the $\Upsilon(4S)$ resonance, we consider a modification
of the usual quantum-mechanical time evolution with a dissipative
term, which contains only one parameter denoted by $\lambda$ and
respects complete positivity. In this way a decoherence effect is
introduced in the time evolution of the 2-particle $B^0 \bar B^0$
state, which becomes stronger with increasing distance between the two
particles. While our model of time evolution has decoherence for the
2-particle system, we assume that, after
the decay of one of the two B mesons, the resulting 1-particle state
obeys the purely quantum-mechanical time evolution. From the
data on dilepton events 
we derive an upper bound on $\lambda$. We also show how $\lambda$ is related
to the so-called ``decoherence parameter'' $\zeta$, which
parameterizes decoherence in neutral flavoured meson--antimeson systems.
\end{abstract}


\section{The introduction}

There is increasing interest in the last years in using particle
physics phenomena for the study of possible
deviations from quantum mechanics (QM). Efforts have been concentrated on
two types of phenomena: oscillations, like $K^0$-$\bar K^0$
\cite{ellisHNS} and neutrino oscillations \cite{lisi}, and quantum
entanglement, where particularly 
suitable systems are the entangled $K^0 \bar K^0$ and $B^0 \bar B^0$
states \cite{six} which are produced in $e^+ e^-$ collisions at the resonances
$\Phi$ and $\Upsilon(4S)$, respectively. These states become
macroscopically extended objects before they decay. Thus in both types
of phenomena macroscopic distances are involved. Furthermore, entangled
systems are -- due to EPR-Bell correlations \cite{bell} -- 
important objects to clearly test QM against local
realistic theories. Whereas entangled systems like $K^0 \bar K^0$ have
a rather longstanding and venerable position in the QM literature
\cite{venerable}, the physics of neutrino oscillations is a rather recent
testing ground for QM; this development has been boosted, in
particular, by the now well-established atmospheric neutrino anomaly,
but also solar neutrinos and neutrinos in the early universe are 
discussed in this context.

In this paper we concentrate on possible decoherence effects which
might arise due to some fundamental modification of QM or
due to the interaction of the system with its ``environment'', whatever
this may be. In the latter case, the idea of the influence of quantum 
gravity \cite{hawking,hooft} -- quantum fluctuations in the
space-time structure at the Planck mass scale -- is especially
attractive nowadays. Possible effects of the environment have 
been investigated intensively in the $K^0 \bar K^0$ system in 
Refs.~\cite{ellisHNS,ellisLMN,banksSP,huetP,benatti97,benatti99,%
benatti01,andrianov}. 
But also other models of
decoherence, like those found in
Refs.~\cite{GRW,pearle,gisinP,penrose}, 
may serve as working hypothesis.
Our model, which we will propose in this paper, has some
remote similarity with the models mentioned here, but ours will be
tailored to the situation of two particles moving apart 
in their center of mass system.

In the past, a measure of decoherence for entangled systems has been
introduced on pure phenomenological grounds in order to determine
quantitatively deviations from pure QM. This simple
procedure of multiplying the quantum-mechanical interference term 
by $1 - \zeta$ \cite{eberhard}, where $\zeta$ is called decoherence
parameter, is a basis-dependent concept and works very well 
as a measure for interpolating continuously between pure QM 
($\zeta = 0$) and total decoherence ($\zeta = 1$). The latter case 
corresponds to spontaneous factorization, also called 
Furry's hypothesis \cite{furry}.
By investigating certain observables, 
the authors of Refs.~\cite{BG97,dass,BG98,BGH99,hiesmayr}
could show that the entangled $K^0 \bar K^0$ and $B^0 \bar B^0$
systems are far from total decoherence, at least when $\zeta$ is
introduced in relation to the basis of mass eigenstates, so that local
realistic theories are highly unlikely. 
In other words, the presence of the interference term is well
established in agreement with QM (see also Ref.~\cite{CPLEAR98}), 
which means that there is quantum interference
of massive particles over macroscopic distances.

In this paper we want to present a model of dissipation for entangled
systems of two particles. In contrast to the prevailing concept in the
literature, where dissipation is introduced at the 1-particle level
and transferred to the 2-particle level through the tensor product structure
of the Hilbert space of states (see, e.g., Ref.~\cite{benatti99}), 
we assume the usual
quantum-mechanical time evolution for the 1-particle states. Thereby
we have in mind that entangled 2-particle systems become decoherent
when they move apart over macroscopic distances, whereas for a
1-particle system QM is not modified. Our dissipative term in the
2-particle time evolution obeys the condition of complete positivity
\cite{lindblad}.
For reasons given below, we consider the entangled $B^0 \bar B^0$ system
with negative C parity. 
By using the experimental value of the ratio $R$ of the number
of like-sign dilepton events over opposite-sign dilepton events, we
can derive a bound on the strength of the dissipative term. 
Considering the time-integrated dilepton event rates,
our model reproduces precisely the corresponding calculations with the
phenomenological decoherence parameter $\zeta$ associated with the
$B_H$-$B_L$ basis. As a result, we obtain a remarkably simple
formula which relates the dissipative strength to the decoherence
parameter $\zeta$. In the context of the observable $R$, we also
compare our model of 2-particle decoherence 
with the case where the analogous dissipative term is introduced already at
the 1-particle level.

\section{The model}

Before considering the $B^0 \bar B^0$ system, let us first discuss our model
of decoherence in a 2-dimensional Hilbert space of states
$\mathcal{H} = \mathbf{C}^2$. We allow for a non-hermitian Hamiltonian $H$, in
order to include the possibility of incorporating particle decay in the
Weisskopf--Wigner approximation \cite{WWA}. We denote the normalized
energy eigenstates
by $|e_j\rangle$ ($j=1,2$) and have, therefore,
\begin{equation}
H |e_j\rangle = \lambda_j |e_j\rangle
\quad \mbox{with} \quad
\lambda_j = m_j - \frac{i}{2} \Gamma_j \,,
\end{equation}
where $m_j$ and $\Gamma_j$ are real and the latter quantities are positive in
addition. Furthermore, we make the crucial assumption that
\begin{equation}
\langle e_1 | e_2 \rangle = 0
\end{equation}
despite the non-hermiticity of $H$.
Including decoherence, the time evolution of the density matrix $\rho$ has the
form
\begin{equation}\label{timeev}
\frac{d \rho}{dt} = -i H \rho + i \rho H^\dagger - D[ \rho ] \,.
\end{equation}
Our model of decoherence consists in assuming that
\begin{equation}\label{D}
D[ \rho ] = \lambda \left( P_1 \rho P_2 + P_2 \rho P_1 \right) \,,
\quad \mbox{where} \quad P_j = | e_j \rangle \langle e_j |
\end{equation}
and $\lambda$ is a positive constant. Such a term is also employed,
for instance, in the context of neutrinos in the early universe (see,
e.g., Ref.~\cite{enqvist}). It can readily be checked that the
decoherence term in Eq.~(\ref{D}) is of the Lindblad type \cite{lindblad}
\begin{equation}\label{compos}
D[ \rho ] = \frac{1}{2}
\left( \sum_j A_j^\dagger A_j \rho + \rho \sum_j A_j^\dagger A_j \right) -
\sum_j A_j \rho A_j^\dagger \,,
\end{equation}
if we make the identification $A_j = \sqrt{\lambda} P_j$. Thus, the term
(\ref{D}) generates a completely positive map; moreover, since
$P_j^\dagger = P_j$ and $[P_j,H]=0$, the decoherence term would increase the
``von Neumann entropy'' and conserve energy in the case of a hermitian
Hamiltonian (see Ref.~\cite{adler} and references therein). However, what is
more important in our discussion is the fact that with the choice (\ref{D})
the equations for the components of $\rho$ decouple. Indeed, with
\begin{equation}
\rho = \sum_{j,k=1}^2 \rho_{jk} | e_j \rangle \langle e_k | \,,
\end{equation}
where $\rho_{jk} = \rho_{kj}^*$,
and with the time evolution (\ref{timeev}), we obtain
\begin{equation}\label{rho(t)}
\begin{array}{l}
\rho_{11}(t) = \rho_{11}(0) \exp (-\Gamma_1t), \\
\rho_{22}(t) = \rho_{22}(0) \exp (-\Gamma_2t), \\
\rho_{12}(t) = \rho_{12}(0) \exp \left\{- [i(m_1 - m_2) +
(\Gamma_1 + \Gamma_2)/2 + \lambda ]\, t \right\} \,.
\end{array}
\end{equation}

Let us for a moment dwell on the motivation
for our model of decoherence. To this end we start with the more general
setting $A_j = \sqrt{\lambda_j} P_j$ ($j=1,2$ and $\lambda_j > 0$) with 
$P_j =| p_j \rangle \langle p_j |$, where the normalized vectors 
$| p_j \rangle$  are linearly independent and in general different from the
eigenvectors of the Hamiltonian. Note that we also allow for
$\langle p_1 | p_2 \rangle \neq 0$. In any case,
one can use formula (\ref{compos}) to obtain a
completely positive dissipative term in the time evolution
(\ref{timeev}), but in general one will not obtain the form of $D[\rho]$ given
by Eq.~(\ref{D}). For the time being
we want to assume that $H^\dagger = H \neq 0$ holds and that $H$ is
non-degenerate in order to avoid trivial considerations.
Now we have two
possibilities: 1.\ The system
$\{ | p_1 \rangle, | p_2 \rangle \}$ is the system of eigenvectors
of $H$, i.e., it is equivalent to the orthonormal system
$\{ | e_1 \rangle, | e_2 \rangle \}$;
2.\ $\{| p_1 \rangle$, $| p_2 \rangle \}$ is not equivalent to the system of
eigenvectors of $H$. In the first case one can show that the form (\ref{D}) of
$D[\rho]$ with $\lambda = (\lambda_1 + \lambda_2)/2$ is obtained and that
\begin{equation}\label{case1}
\mbox{Case 1}\: \Leftrightarrow\: [H,P_j] = 0 \; \mbox{ for }\;
j=1,2 \,: \quad
\lim_{t \to \infty} \rho(t) = P_1 \rho(0) P_1 + P_2 \rho(0) P_2 
\end{equation}
holds.
Furthermore, density matrices $P_j$, or linear combinations
thereof, are constant solutions of the time evolution equation.
In the second case, at least one of the vectors 
$| p_1 \rangle$, $| p_2 \rangle$ is not an eigenvector of $H$ and
one can prove that
\begin{equation}\label{case2}
\mbox{Case 2}\: \Leftrightarrow\: 
\exists\, j = 1 \:\mbox{or}\: 2 \;\mbox{ with }\; [H,P_j] \neq 0\,: \quad
\lim_{t \to \infty} \rho(t) = {\textstyle \frac{1}{2}} \mathbf{1} \,,
\end{equation}
independent of $\rho(0)$.
We will see in the following -- when we apply our model to the
$B^0 \bar B^0$ system -- that the first case is closer to our physical
intuition (see also last paragraph of this section). 
The two cases have been described in
Refs.~\cite{chang,benatti00} in the context of 1-particle decoherence
in neutrino oscillations. Note that Case 1 is used in Ref.~\cite{lisi}
(see also Ref.~\cite{adler}),
whereas Case 2 is considered, e.g., in Ref.~\cite{klapdor}
in the same context.
If we allow for $H^\dagger \neq H$, the picture, we have
developed here, gets blurred because then there is a competition between
particle decay, i.e., $\lim_{t \to \infty} \mbox{Tr}\, \rho(t) = 0$, and
the effect of decoherence. We nevertheless stick to the first
case. Note that identifying the orthonormal system given by $P_j$
($j=1,2$) with the system of eigenvectors of $H$ is not only
motivated by the considerations above but also by simplicity; as we
have seen in Eq.~(\ref{rho(t)}) we have decoupled time evolutions as a
consequence. We want to stress, however, that in the case of CP
violation, which is particularly important for the $K^0 \bar K^0$
system, it might be useful to allow for small deviations from
$\langle e_1 | e_2 \rangle = 0$, and thus for small deviations of the
eigenvectors of the Hamiltonian from the orthonormal system
$\{ | p_1 \rangle, | p_2 \rangle \}$.

Returning from general considerations, we now
we apply our model of decoherence to the case of the 2-particle
$B^0 \bar B^0$ state, generated by the decay of the $\Upsilon(4S)$
resonance (for the formalism used in the $B^0 \bar B^0$ system see, e.g.,
Ref.~\cite{luis}). We conceive $t$ as the eigentime of $B^0$ and 
$\bar B^0$ and make the identification
\begin{equation}\label{e1e2}
| e_1 \rangle = | B_H \otimes B_L \rangle
\quad \mbox{and} \quad
| e_2 \rangle = | B_L \otimes B_H \rangle \,,
\end{equation}
where the heavy and the light neutral $B$ states are defined via
\begin{equation}
| B_H \rangle = p | B^0 \rangle + q | \bar B^0 \rangle
\quad \mbox{and} \quad
| B_L \rangle = p | B^0 \rangle - q | \bar B^0 \rangle \,,
\end{equation}
which have eigenvalues
\begin{equation}
\lambda_H = m_H - \frac{i}{2} \Gamma_H
\quad \mbox{and} \quad
\lambda_L = m_L - \frac{i}{2} \Gamma_L \,,
\end{equation}
respectively, of the effective 1-particle Hamiltonian $H_1$. For the
2-particle system, we transfer, as usual, the 1-particle Hamiltonian
to the tensor product of the 1-particle Hilbert spaces by using
$H = H_1 \otimes \mathbf{1} + \mathbf{1} \otimes H_1$.
We imagine that the first factor in the tensor product corresponds to
particles moving to the left,
whereas the second factor in the tensor product corresponds to
right-moving particles.
We assume CP conservation in $B^0$-$\bar B^0$ mixing, which is a
good approximation \cite{luis,RPP} and corresponds to $|p/q| = 1$.
In this case we have $\langle B_H | B_L \rangle = 0$ and, therefore,
$\langle e_1 | e_2 \rangle = 0$. In the following we will
set $p = q = 1/\sqrt{2}$.
At the $\Upsilon(4S)$ resonance, at $t=0$, the entangled state
\begin{equation}\label{psi}
| \psi \rangle =
\frac{1}{\sqrt{2}} \left( | e_1 \rangle - | e_2 \rangle \right)
\end{equation}
is produced, which is equivalent to the density matrix
\begin{equation}\label{rho(0)}
\rho(0) = \frac{1}{2} \left(
| e_1 \rangle \langle e_1 | + | e_2 \rangle \langle e_2 | -
| e_1 \rangle \langle e_2 | - | e_2 \rangle \langle e_1 | \right)\,.
\end{equation}
With the time evolution (\ref{rho(t)}), the initial condition (\ref{rho(0)})
and taking into account that in the case of
the vectors (\ref{e1e2}) we have
$\lambda_1 = \lambda_2 = m_H + m_L - i \Gamma$ with
$\Gamma \equiv (\Gamma_H + \Gamma_L)/2$, we obtain the time evolution
\begin{equation}\label{rho}
\rho(t) = \frac{1}{2} e^{-2\Gamma t} \left\{
| e_1 \rangle \langle e_1 | + | e_2 \rangle \langle e_2 | -
e^{-\lambda t} \left(
| e_1 \rangle \langle e_2 | + | e_2 \rangle \langle e_1 |
\right) \right\} \,.
\end{equation}
Note that the factor $\exp (-\lambda t)$ in the density matrix (\ref{rho})
introduces decoherence as a consequence of the $D$-term in the time
evolution (\ref{timeev}). In other words, for $t>0$ and $\lambda > 0$,
the density matrix (\ref{rho}) does not correspond to a pure state
anymore.

Having chosen the energy eigenstates (\ref{e1e2}) for the construction
of the projectors $P_j$, our model complies with Case 1
(\ref{case1}). In this case, we would have no decoherence if
$\rho(0) = P_1$ or $P_2$, though such initial conditions might be
unrealistic. This agrees with our intention because in these
cases we have no entanglement over macroscopic distances and no reason
for modifying the quantum-mechanical time evolution.
Note that using the projector states (\ref{e1e2}) confines our
Hilbert space of states to a 2-dimensional one. Using projector states
which are non-trivial orthogonal linear combinations of the states
(\ref{e1e2}), would lead to Case 2 (\ref{case2}),
still with a 2-dimensional Hilbert space.
However, using projector states
which are \emph{not} linear combinations of the states (\ref{e1e2}),
like, e.g.,
$| B^0 \otimes \bar B^0 \rangle$, $| \bar B^0 \otimes B^0 \rangle$,
entails not only a time evolution into the full 4-dimensional Hilbert
space of states, including
$| B_H \otimes B_H \rangle$ and $| B_L \otimes B_L \rangle$,
but also opens up the possibility for more involved schemes of
decoherence than given by our simple model.

\section{The measurement}

In order to obtain information on the parameter $\lambda$,
which modifies the time
evolution in the $B^0 \bar B^0$ system, we adopt the following
philosophy. We start at $t=0$ with the density matrix (\ref{rho(0)})
for a $B^0 \bar B^0$ state with negative C parity. This 2-particle
density matrix follows the time evolution (\ref{rho}) and undergoes
thereby some decoherence. We imagine a measurement of the $B$ quantum
number of the left-moving particle at time $t_\ell$ and of the
right-moving particle at time $t_r$. For times
$\min (t_\ell,t_r) < t < \max (t_\ell,t_r)$ we have a 1-particle state
which we assume to evolve exactly according to QM, with
the time evolution given by $H_1$. 

In a mathematical language, we do the following. Assuming for
definiteness $t_\ell < t_r$, at $t=t_\ell$ we calculate the trace
\begin{equation}
\mathrm{Tr}_\ell \left\{
\left( | n \rangle \langle n | \otimes \mathbf{1} \right) \rho(t_\ell)
\right\} \equiv \rho_r(t_\ell;t=t_\ell)\,,
\end{equation}
where $\mathrm{Tr}_\ell$ means the trace evaluated only in the space
of the left-moving particles; $\rho(t_\ell)$ is given by
Eq.~(\ref{rho}), evaluated at $t = t_\ell$; moreover, we have defined
$| 1 \rangle = | B^0 \rangle$ and $| 2 \rangle = | \bar B^0 \rangle$,
and $n = 1,2$. Consequently,
$\rho_r(t_\ell;t=t_\ell)$ is a 1-particle density matrix for the right-moving
ones. For $t>t_\ell$, it is denoted by $\rho_r(t_\ell;t)$ and
follows the 1-particle time evolution. At
$t = t_r$, where we measure the $B$ quantum number of the right-moving
particles, we finally have ($n' = 1,2$)
\begin{equation}
N(n,t_\ell;n',t_r) = \mathrm{Tr}\, \left\{
| n' \rangle \langle n' | \rho_r(t_\ell;t_r) \right\}\,.
\end{equation}

Using all the above formalism and allowing also for $t_\ell > t_r$,
we arrive at
\begin{eqnarray}
\lefteqn{N(n,t_\ell;n',t_r) = \frac{1}{2} e^{-\Gamma (t_\ell + t_r)}}
\nonumber \\
&& \times \left\{
| \langle n | B_H \rangle |^2\, | \langle n' | B_L \rangle |^2 \,
e^{-\Delta \Gamma (t_\ell - t_r)/2} +
| \langle n | B_L \rangle |^2\, | \langle n' | B_H \rangle |^2 \,
e^{\Delta \Gamma (t_\ell - t_r)/2}
\right. \nonumber \\
&& -\, e^{-\lambda \min (t_\ell,t_r)} \left(
\langle n | B_H \rangle \langle n | B_L \rangle^*
\langle n' | B_L \rangle \langle n' | B_H  \rangle^*
e^{-i\Delta m (t_\ell - t_r)} \right. \nonumber \\
&& \left. \left. \hphantom{xxxxxxxxl\,} +
\langle n | B_L \rangle \langle n | B_H \rangle^*
\langle n' | B_H \rangle \langle n' | B_L  \rangle^*
e^{i\Delta m (t_\ell - t_r)}
\right) \right\}\,. \label{N}
\end{eqnarray}
In this equation we have used the notation
$\Delta \Gamma = \Gamma_H - \Gamma_L$ and $\Delta m = m_H - m_L$.
For the sake of clarity, we have retained the scalar products
$\langle n | B_{H,L} \rangle$ and $ \langle n' | B_{H,L}
\rangle$. According to our assumption of CP conservation in
$B^0$-$\bar B^0$ mixing, we will replace them by their values
$\pm 1/\sqrt{2}$. It is easy to check that for $\lambda = 0$ one
obtains the usual expressions found in the literature.
Note that for $t_\ell = t_r$ and $n = n'$ we have
\begin{equation}
N(n,t_\ell;n,t_\ell) = \frac{1}{4} e^{-2\Gamma t_\ell}
\left( 1 - e^{-\lambda t_\ell} \right) \,,
\end{equation}
which is different from zero, in contrast to the standard
quantum-mechanical case.

\section{The dileptonic decays}

In practice, measurement of the $B$ quantum number of neutral mesons in
the entangled $B^0 \bar B^0$ state proceeds via flavour tagging when
the mesons decay.
Assuming the validity of the $\Delta B = \Delta Q$ rule, in inclusive
semileptonic decays $\ell^+$ tags $B^0$ and $\ell^-$ tags $\bar B^0$
($\ell = e$ or $\mu$). In the following we will concentrate on
dilepton events \cite{carter}. 
Denoting the inclusive semileptonic decay rate by
$\Gamma_\ell$, the numbers of dilepton events
from the decay of $| \psi \rangle$ (\ref{psi}) are then
given by the integrals
\begin{eqnarray}
&& N_{++} = \Gamma_\ell^2 \int_0^\infty dt_\ell \int_0^\infty dt_r
\, N(1,t_\ell\,;1,t_r)\,, \nonumber \\
&& N_{--} = \Gamma_\ell^2 \int_0^\infty dt_\ell \int_0^\infty dt_r
\, N(2,t_\ell\,;2,t_r)\,, \nonumber \\
&& N_{+-} = N_{-+} = \Gamma_\ell^2 \int_0^\infty dt_\ell \int_0^\infty dt_r
\, N(1,t_\ell\,;2,t_r)\,.
\end{eqnarray}
Defining $x = \Delta m/\Gamma$ and $y = \Delta \Gamma/2\Gamma$ and
calculating these integrals leads to the result
\begin{eqnarray}
&& N_{++} = N_{--} = \frac{\Gamma_\ell^2}{4 \Gamma^2}
\left\{ \frac{1}{1-y^2} - \frac{1}{1+x^2}
\left( 1 - \zeta(\Lambda) \right) \right\}\,, \label{N++} \\
&& N_{+-} = N_{-+} = \frac{\Gamma_\ell^2}{4 \Gamma^2}
\left\{ \frac{1}{1-y^2} + \frac{1}{1+x^2}
\left( 1 - \zeta(\Lambda) \right) \right\}\,, \label{N+-}
\end{eqnarray}
where the function $\zeta(\Lambda)$ is given by the simple expression
\begin{equation}\label{zeta}
\zeta(\Lambda) = \frac{\Lambda}{2+\Lambda}
\quad \mbox{with} \quad \Lambda = \frac{\lambda}{\Gamma}\,.
\end{equation}
It is interesting to note that $x$ does not enter into $\zeta$.

Eqs.~(\ref{N++}) and (\ref{N+-}) reproduce the results of
Refs.~\cite{dass,BG98}, where the ``decoherence parameter'' $\zeta$
\cite{eberhard} is introduced phenomenologically in the observable 
\begin{equation}\label{R}
R = \frac{N_{++} + N_{--}}{N_{+-} + N_{-+}} \,,
\end{equation}
by multiplying the interference terms in $N_{++}$, etc., with
$(1-\zeta)$. In the model presented here, $\zeta$ is expressed by the
parameter $\lambda$ (see Eq.~(\ref{zeta})), the strength of the
dissipative term in the modified time evolution (\ref{timeev}).
It has been discussed in the literature that the above phenomenological
procedure of introducing a parameter $\zeta$ depends on the basis
chosen in $B^0$-$\bar B^0$ space \cite{furry,dass,BG98,BGH99,DG00}.
In the present model it is the $\zeta$ associated with the
$B_H$-$B_L$ basis.

Let us perform a numerical estimate of $\zeta$ and $\Lambda$ along the
lines presented in Ref.~\cite{BG97}.
To this end we use $R$ (\ref{R}),
which has been measured by the ARGUS \cite{ARGUS} and CLEO \cite{CLEO}
Collaborations. Combining both measurements, we obtain the value
$R_\mathrm{exp} = 0.189 \pm 0.044$ \cite{BG97}. As far as $x$ is
concerned, we use the value $x_\mathrm{exp} = 0.740 \pm 0.031$
obtained by combining the data from all LEP experiments \cite{RPP}.
With the approximation $y = 0$ in Eqs.~(\ref{N++})
and (\ref{N+-}) \cite{RPP,buras} and using the law of propagation of
errors, from $R_\mathrm{exp}$ and
$x_\mathrm{exp}$ we derive the following numerical estimates:
\begin{equation}
\zeta = -0.06 \pm 0.10
\quad \mbox{and} \quad
\Lambda = -0.11 \pm 0.18.
\end{equation}
The Belle Collaboration has published data on the correlated semileptonic
decay rate as a function of the difference $t_\ell - t_r$ \cite{Belle}. Of
course, these data could also be used to put a limit on $\lambda$, if we
integrate $N(n,t_\ell;n',t_r)$, Eq.~(\ref{N}), over $t_\ell + t_r$. However, we
do not have enough information to perform such a fit.

Let us now compare our model, where decoherence is implemented at the
2-particle level, with the case where we have the analogous time
evolution (\ref{timeev}) at the 1-particle level
\cite{benatti99,benatti01}. We use the same 
structure of the $D$-term as given by Eq.~(\ref{D}), but now with
$| e_1 \rangle = | B_H \rangle$ and $| e_2 \rangle = | B_L \rangle$,
instead of Eq.~(\ref{e1e2}). We denote the strength of the
dissipative term by $\xi$, in order to distinguish it from $\lambda$
in the case of 2-particle decoherence. Evidently, we have the same time
evolution (\ref{rho(t)}) at the 1-particle level, with $\lambda$
replaced by $\xi$. Following the steps to derive
Eq.~(\ref{N}), we obtain the same formula, except that
$\exp (- \lambda \min (t_\ell,t_r))$ is replaced by
$\exp (- \xi (t_\ell + t_r))$. Eventually, we arrive at $N_{++}$ and
$N_{+-}$ given by Eqs.~(\ref{N++}) and (\ref{N+-}), respectively, where
$\zeta$ is now given by\footnote{Note that the result for
$R$ with $\zeta$ given by Eq.~(\ref{zetaxi}) agrees with the result
for $R$ in first order in $\Xi$ of Ref.~\cite{benatti99}, if the
general Lindblad term in this paper is specialized to energy conservation.
We thank the referee for pointing this out to us.}
\begin{equation}\label{zetaxi}
\zeta(\Xi,x) = \frac{(1+\Xi)^2 - 1}{(1+\Xi)^2 + x^2}
\quad \mbox{with} \quad
\Xi = \frac{\xi}{\Gamma}\,.
\end{equation}
Thus the two models of decoherence cannot be distinguished on the
basis of the time-integrated dilepton event rates, but only on the
basis the time-dependent event rates.
A numerical estimate analogous to the one performed for $\Lambda$
leads to the result $\Xi = -0.04 \pm 0.07$. 

\section{The summary}

In this paper we have considered a model of decoherence applicable in the
center of mass system of two particles. Our model reflects the idea that, when
the two particles move apart and eventually become macroscopically separated,
some ``forces'' might be operative which de-entangle the quantum-mechanical
state as a function of the distance. The dissipative term which we have added
to the quantum-mechanical time evolution could be an effective term
originating in some modification of QM; it could as well be
based on some effective quantum-mechanical description of an interaction of
the 2-particle system with an unknown environment. Our dissipative term
respects complete positivity, which -- we believe -- is a useful physical
guiding principle for modifications of the quantum-mechanical time evolution.
In compliance with our idea, we assume that, after one of the particles has
decayed, the other one follows the quantum-mechanical time evolution. We have
applied our model in the case of the entangled $B^0 \bar B^0$ state with
negative C parity, where we have used the data on the B lifetime,
the $B_H$-$B_L$ mass difference measured by observing the time evolution of
single neutral B mesons, and the ratio of like-sign over opposite-sign
dilepton event rates for the purpose of estimating the strength
$\lambda$ of the dissipative term.
In the case of time-integrated dilepton events, our simple model
leads to a result which is also obtained by the phenomenological
introduction of a ``decoherence parameter'' $\zeta$
in the quantum-mechanical interference terms of the
quantities $N_{++}$, $N_{+-}$, etc. In the dissipative term $D[\rho]$ we
employ the states
$| B_H \otimes B_L \rangle$ and $| B_L \otimes B_H \rangle$;
this, eventually, modifies the interference
terms of $N_{++}$, $N_{+-}$, etc., with the $\zeta$ associated with the
$B_H$-$B_L$ basis \cite{dass,BG98}.
Note that we have neglected CP violation in $B^0$-$\bar B^0$
mixing, which is a good approximation in this system. Transferring our model
of decoherence to the $K^0 \bar K^0$ system is
not straightforward, because it requires to take CP violation and the
non-orthogonality of the $K_S$ and $K_L$ states into account. Work on this is
in progress.

\end{document}